\documentclass[twocolumn,showpacs,preprintnumbers,amsmath,amssymb]{revtex4}
\usepackage{graphicx}% Include figure files
\usepackage{dcolumn}% Align table columns on decimal point
\usepackage{bm}% bold math
\usepackage{color}
\newcommand{\sgm}{$\sigma_{\rm s}$}
\newcommand{\msr}{$\mu$SR}
\usepackage{color}

\begin{document}
\preprint{APS/123-QED}
\title{Anisotropic order parameter\\
in Li-intercalated layered superconductor Li\bm{$_x$}ZrNCl}
\author{M.~Hiraishi$^1$}
\author{R.~Kadono$^{1,2}$}
\author{M.~Miyazaki$^1$}
\author{S.~Takeshita$^2$}
\author{Y.~Taguchi$^3$}
\author{Y.~Kasahara$^4$}
\author{T.~Takano$^4$}
\author{T.~Kishiume$^4$}
\author{Y.~Iwasa$^4$}
\affiliation{
\vspace{2mm}
$^{\rm 1}$Department of Materials Structure Science, The Graduate University for Advanced Studies, Tsukuba, Ibaraki 305-0801, Japan\\
$^{\rm 2}$Institute of Materials Structure Science, High Energy Accelerator Research Organization, Tsukuba, Ibaraki 305-0801, Japan\\
$^{\rm 3}$Cross-correlated Materials Research Group(CMRG), ASI, RIKEN, Wako 351-0198, Japan\\
$^{\rm 4}$Institute for Materials Research, Tohoku University, Sendai 980-8577, Japan
}%

\date{\today}
\begin{abstract}
In this study, it is shown that in a layered nitride superconductor, {\it i.e.}, Li$_x$ZrNCl ($0.07\le x\le0.21$), the superconducting order parameter is highly anisotropic in a sample with $x=0.12$, as inferred from both the temperature and the magnetic field dependences of the muon depolarization rate ($\sigma_{\rm s}$, proportional to the superfluid density). Moreover, the tendency of strong anisotropy with an increase in $x$ is indicated by the $T$ dependence of \sgm. These observations are in good agreement with the recent theory that predicts the development of anisotropy in a $d+id'$ gap upon carrier filling to the bands with disconnected Fermi surfaces on a honeycomb lattice.
\end{abstract}

\maketitle
Layered nitrides such as $\beta$-$M$NCl (with $M=$ Zr, Hf) are attracting considerable attention since they exhibit superconductivity upon intercalating alkaline metals \cite{Yamanaka:96,Yamanaka:98}. While these nitrides have relatively high superconducting transition temperatures [$T_{\rm c}\simeq15$ (25)~K with $M=$ Zr (Hf)], the density of states at the Fermi level [$N(0)$] is reported to be considerably lower than that of other superconductors having similar $T_{\rm c}$, as inferred from the results of the specific heat measurement of Li$_{0.12}$ZrNCl \cite{Taguchi:05} and magnetic susceptibility measurement of Li$_{0.48}$(THF)$_y$HfNCl (where THF refers to tetrahydrofuran) \cite{Tou:01}. The small $N(0)$ is in excellent agreement with the prediction of the theoretical investigation\cite{Weht}. Such situation naturally leads to the question of whether or not the superconductivity in layered nitrides is fully understood on the basis of the conventional BCS theory with electron-phonon coupling. Moreover, it has been shown that one of these nitrides, {\it i.e.}, Li$_x$ZrNCl, exhibits further anomalies that are not expected to be present in the simplest situation presumed by the BCS theory.

Li$_x$ZrNCl has a lamellar structure consisting of alternating stacks of Zr-N double honeycomb layers and insulating Cl bilayers. Li atoms are intercalated into the van der Waals gap of the Cl bilayers to supply electrons to the conducting Zr-N layers. In addition to the above mentioned small $N(0)$, it has been also revealed that the electron-phonon interaction is too weak to explain its high $T_{\rm c}$\cite{Taguchi:05,Weht,Heid}. In general, in the case of two-dimensional electronic systems such as $\beta-$ZrNCl, $N(0)$ may be only weakly dependent on band filling. Therefore, provided that superconductivity is explained by the conventional BCS theory, $T_{\rm c}$ would not vary with the Li concentration ($x$). However, the fact is that while $T_{\rm c}$ is independent of doping for $x\geq0.15$, it increases steeply below $x\simeq0.12$, reaching a maximum ($T_{\rm c}=15.2$~K at $x=0.06$) and then suddenly transitioning into an insulating state for $x\le0.05$ \cite{Taguchi:06}. It might be worth noting that this tendency of $T_{\rm c}$ to be high at a low carrier density $x$ is opposite to that of underdoped cuprates. 

Another anomaly is reported in the magnetic field dependence of an electronic specific heat (Sommerfeld) coefficient $\gamma$ in the mixed state of Li$_{0.12}$ZrNCl. While $\gamma$ is expected to be approximately proportional to the number of flux lines, and accordingly to the magnetic field in conventional BCS superconductors ({\it i.e.}, $\gamma\simeq\gamma_nH/H_{c2}\propto H/\Phi_0$, where $\gamma_n$ is the electronic specific heat in the normal state, $H$ is the magnetic field, $H_{c2}$ is the upper critical field, and $\Phi_0=2.07\times10^{-15}$~T$\cdot$m$^2$ is the quantum flux), it increases with a gradient much steeper than $\gamma_n/H_{c2}$ in Li$_{0.12}$ZrNCl, approaching $\gamma_n$ at $H\sim0.4H_{c2}$ \cite{Taguchi:05}. This strongly suggests the occurrence of field-induced quasiparticle excitation that is not expected for the superconducting order parameter described by isotropic $s$-wave paring with a unique gap energy.

In this paper, we describe our muon spin rotation (\msr) study on Li$_x$ZrNCl over a range of Li content near metal-to-insulator transition ($0.07\le x\le0.21$). We show that in a sample with $x=0.12$, the temperature and magnetic field dependences of the muon spin depolarization rate [$\sigma_{\rm s}\equiv\sigma_{\rm s}(T,H)$, which is proportional to the superfluid density $n_s$] is in complete agreement with the anomalies observed in the bulk properties, providing microscopic evidence for the anisotropic order parameter.
Furthermore, a clear tendency of increasing anisotropy as a function of $x$ is inferred from the magnitude of the gap ratio ($2\Delta/k_{\rm B}T_{\rm c}$). These observations support the recent theory that predicts the occurrence of anisotropic $d+id'$ pairing in Li$_x$ZrNCl and the effect of electronic correlations on the Fermi surface specific to Li$_x$ZrNCl, which develops with band filling \cite{kuroki:08,kuroki:09}.

\begin{figure}[t]
	\centering
	\includegraphics[width=1.0\linewidth,clip]{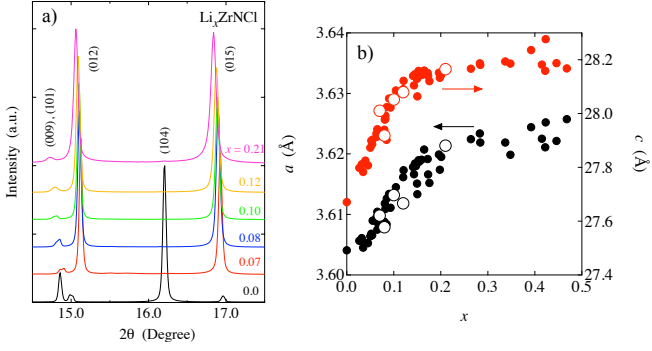}
	\caption{(Color online) (a) X-ray diffraction spectra for the present samples obtained
	at SPring-8 (X-ray energy of 15.6 keV). The [104] line observed in the pristine compound is absent in 
	Li-doped samples. (b) Lattice parameters (open circles show present data, 
	and filled points are after Ref.\cite{Taguchi:06}).} \label{xray}
\end{figure}

Conventional \msr\ measurements were performed on M15 and M20 beamlines of TRIUMF, Canada. A \msr\ apparatus with high time resolution was used to measure the time-dependent positron decay asymmetry under a transverse field (TF) up to 2~T. Li$_x$ZrNCl samples, alined along the $c$ axis, were loaded on a He gas-flow cryostat, and they were field cooled to a target temperature to minimize the effect of flux pinning. Details of sample preparation are described in the earlier report, where special precaution was taken to maintain the homogeneity of samples \cite{Taguchi:06}. The superconducting volume fraction as well as $T_{\rm c}$ [$= 15.1$~K ($x=0.07$), 14.0~K ($x=0.08$), 12.6~K ($x=0.10$), 12.5~K ($x=0.12$) and 11.6~K ($x=0.21$)] determined from magnetization measurements were in excellent agreement with the previous result \cite{Taguchi:06}. The concentration of lithium was determined from inductively coupled plasma spectroscopy. As shown in Fig.~\ref{xray}, the homogeneity of samples was confirmed by carrying out high-resolution powder X-ray diffraction at SPring-8. The mean free path $l$ of these samples were estimated to be 10--16 nm (increasing with $x$)\cite{Takano:08}, which is approximately equal to the coherence length ($\xi=$ 8--13 nm). Although this would not be in the clean limit, the condition $l>\xi$ is always satisfied, and thus, the situation is consistent with the observed influence of anisotropic gap on \sgm.

\msr\ is an effective microscopic technique for measuring the magnetic penetration depth ($\lambda$) in type II superconductors. It is reasonably presumed that implanted muons are distributed randomly over the length scale of flux line lattice (FLL), probing local magnetic fields at their respective positions.  Then the transverse muon spin precession signal consists of a random sampling of internal field distribution $B({\bf r})$, such that
\begin{align*}
	P^{\rm v}_x(t)&=\int_{-\infty}^{\infty}\cos(\gamma_{\mu}Bt+\phi)n(B)dB,\\
	n(B)&=\langle\delta[B({\bf r})-B]\rangle_{\bf r},
\end{align*}
where $\gamma_{\mu}=2\pi\times135.53$ MHz/T is the muon gyromagnetic ratio, $n(B)$ is the spectral density for the internal field defined as a spatial average $(\langle\rangle_{\bf r})$ of the delta function, and $\phi$ is the initial phase of rotation. Hence, $n(B)$ can be obtained from the real amplitude of the Fourier transform of the TF-\msr\ signal. In the case of a relatively long magnetic penetration depth ($\lambda\geq300$ nm), the Gaussian distribution is a good approximation for $n(B)$, yielding
\begin{align*}
	P^{\rm v}_x(t)\simeq\exp(-\sigma_{\rm s}^2t^2/2)\cos(\omega_0 t+\phi),
\end{align*}
where $\sigma_{\rm s}$ is obtained from a second moment of the field distribution ($=\gamma_{\mu}\sqrt{\langle[B({\bf r})-B]^2\rangle}$), and $\omega_0=\gamma_{\mu}B_0$ with $B_0\simeq\mu_0H$. Here, provided that the clean limit is achieved, $\lambda$ is related to the superconducting carrier density $n_{\rm s}$ as follows:
\begin{align*}
	\sigma_{\rm s}\propto\frac{1}{\lambda^2}=\frac{n_{\rm s}e^2}{m^*c^2}.
\end{align*}\par

\begin{figure}[t]
	\centering
	\includegraphics[width=0.7\linewidth,clip]{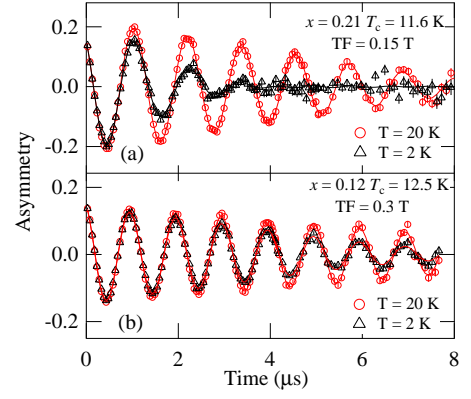}
	\caption{(Color online) Examples of TF-\msr\ spectra with $x=0.21$ (a) and 0.12 (b) shown in a rotating reference frame frequency of 20~MHz and 40~MHz, respectively. Circles represent the spectra above $T_{\rm c}$, while triangles are those obtained below $T_{\rm c}$. Solid lines are fits obtained using Eq.~(\ref{asym}).} \label{tsp}
\end{figure}

Fig.~\ref{tsp} shows some examples of the TF-\msr\ time spectra observed in the samples with $x=0.21$ and 0.12 under a field of 0.15~T and 0.3~T, respectively (shown in a rotating-reference-frame for visibility). While both spectra show a slow Gaussian damping above $T_{\rm c}$ because of random local fields from nuclear magnetic moments, further depolarization resulting from the formation of a flux line lattice is observed upon superconducting transition. A clear trend toward a high depolarization rate is observed for a high Li concentration $x$, which indicates an increase in $n_{\rm s}$ with $x$. We also note that an additional component that exhibits fast depolarization ($\sim$2~MHz) is observed in samples with $x=0.08$, 0.10, and 0.12. This observation is tentatively attributed to muonium (Mu) that may be formed when muons are stopped near the insulating Cl bilayers. Considering a background contribution from muons stopped in the sample holder, we used the following function for the analysis of the \msr\ time spectra by curve fitting:
\begin{align}
	A_0P_x(t)&=\exp(-\sigma_{\rm n}^2t^2)[A_{\rm s}P^{\rm v}_x(t)
	+A_{\rm f}e^{-\Lambda t}\cos(\omega_0t+\phi)]\nonumber\\
	&+A_{\rm b}\exp\left(-\sigma_{\rm b}^2t^2\right)\cos\left(\omega_{\rm b}t+\phi\right),
	\label{asym}
\end{align}
where $A_0$ is the total positron decay asymmetry ($\sim$0.2), $\sigma_{\rm n}$ is the depolarization rate attributed to nuclear magnetic moments, $A_{\rm s}$ is the partial asymmetry of a superconducting fraction, $A_{\rm f}$ is that of the component related with Mu formation showing depolarization at a rate $\Lambda$, $A_{\rm b}$ ($=A_0-A_{\rm s}-A_{\rm f}\le0.01$) and $\omega_{\rm b}$ are the amplitude and central frequency of the background. It was found that $A_{\rm f}$ and $\Lambda$ were mostly independent of temperature and magnetic field, and their fractional yield ($A_{\rm f}/A_0$) was 0.26(1), 0.169(1), and 0.163(1) for $x=0.08$, 0.10, and 0.12, respectively.
The influence of flux pinning on $\sigma_{\rm s}$ was confirmed to be negligible for $x=0.10$ and 0.21, as inferred from the nearly field-independent $\sigma_{\rm s}$ at low magnetic induction ($<0.2$~T). This also confirms the formation of three-dimensional vortices as expected from relatively large coherence length ($\xi=$~8--13~nm) over the Cl-bilayer thickness ($\sim$1~nm).

\begin{figure}[t]
	\centering
	\includegraphics[width=0.7\linewidth,clip]{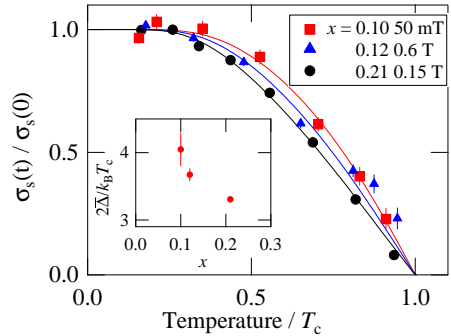}
	\caption{(Color online)Temperature dependence of \sgm\ for samples with $x=0.10$, 0.12, 0.21. Solid curves are fits obtained using the s-BCS theory as an {\sl effective} model. Longitudinal and horizontal axes are normalized by \sgm(0), $T_{\rm c}$ for comparison. Inset shows the $x$ dependence of a fitting parameter $2\overline{\Delta}/k_{\rm B}T_{\rm c}$, with $\overline{\Delta}$ interpreted as a mean value.}		\label{sgmvst}
\end{figure}

Figure \ref{sgmvst} shows a plot of \sgm\ against temperature for the samples in which \sgm\ has been deduced with sufficient precision; it was found that $n_{\rm s}$ was too small for samples with $x<0.10$.  Solid curves are the best fits obtained by applying the conventional BCS theory for $s$-wave symmetry with a single-gap ($s$-BCS theory) \cite{Bouqyuet:01}. Here, assuming that electron-phonon coupling plays a minor role in Li$_x$ZrNCl, we use the $s$-BCS theory to determine the curve fit as an effective model extended to the case of anisotropic order parameters (having dips, nodes, or a secondary gap in multiple bands). This extension is achieved by allowing the gap ratio ($2\Delta/k_{\rm B}T_{\rm c}$) to vary freely, where the energy gap $\Delta$ is regarded as a mean value, $\overline{\Delta}$, averaged over the momentum space and relevant bands (while $T_{\rm c}$ is determined by the maximum of $\Delta$). Then, we find that the deduced gap ratio ($2\overline{\Delta}/k_{\rm B}T_{\rm c}$, shown in the inset of Fig.~\ref{sgmvst}) decreases with an increase in $x$. Assuming that $T_{\rm c}$ {\sl decreases} with an increase in $x$, we find that the decrease in $\overline{\Delta}$ is steeper than that in decrease in $T_{\rm c}$. While this tendency is not easily understood in the framework of electron-phonon coupling, it is understood in the extended model as an indication toward strong anisotropy with high Li content. 

While we cannot distinguish the origin of anisotropy between one-band and multiband scenarios solely from the behavior of superfluid density, the presence of a secondary energy scale in the energy gap is inferred from the analysis using a phenomenological double-gap model for $s$-wave symmetry \cite{Ohishi:01}, such that
\begin{align}
	\sigma_{\rm s}(T)&=\sigma_{\rm s}(0)\left[1-w\delta\sigma(\Delta_1,T)-(1-w)\delta\sigma(\Delta_2,T)\right],\nonumber\\
	\delta\sigma(\delta,T)&=\frac{2}{k_{\rm B}T}\int_0^{\infty}f(\epsilon,T)\cdot[1-f(\epsilon,T)]\,d\epsilon,\nonumber\\
	f(\epsilon,T)&=\left[1+\exp\left(\sqrt{\epsilon^2+\Delta(T)^2}/k_{\rm B}T\right)\right]^{-1}, \nonumber
\end{align}
where $f(\epsilon,T)$ is the Fermi distribution function, $\Delta(T)$ is the standard BCS gap energy, and $w$ is the fractional weight of the $i=1$ component. Although it is not easy to determine all the parameters using the curve fits of data shown in Fig.~\ref{sgmvst}, $\Delta_2$ for $x=0.12$ can be determined on the basis of an assumption that $\Delta_1$ corresponds to $\Delta(T=0)$ determined by a jump in the specific heat near $T_{\rm c}$ (2.64 meV) \cite{Taguchi:05}, yielding excellent fit with the data obtained when $2\Delta_2/k_{\rm B}T_{\rm c}=2.4(4)$ [$\Delta_2=1.3(2)$~meV, or $\Delta_1/\Delta_2=2.0(3)$] and $w=0.59(9)$. 

\begin{figure}[t]
	\centering
	\includegraphics[width=0.7\linewidth,clip]{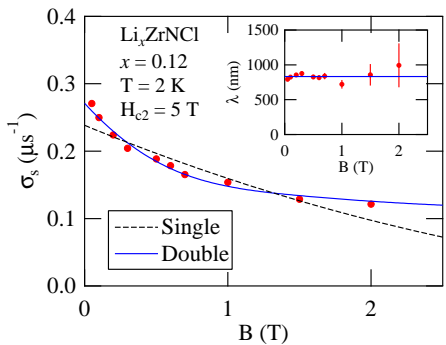}
	\caption{(Color online) Field dependence of \sgm\ in Li$_x$ZrNCl with $x=0.12$. Solid curve represents the best fit obtained by Eq.~(\ref{sgmh}), and the dashed curve corresponds to the case of $z=1$. Inset shows magnetic penetration depth deduced from the same analysis.}\label{sgmvsh}
\end{figure}

The presence of a secondary energy scale in the order parameter is further suggested by the magnetic field dependence of \sgm\ measured at 2~K for the same sample in which an anomalous behavior of $\gamma$ has been reported \cite{Taguchi:05}.
As shown in Fig.~\ref{sgmvsh}, it is observed that in the limit of $H/H_{c2}(\equiv h)\rightarrow1$, \sgm\ exhibits a trend of asymptotic conversion to a value ($\sigma_{{\rm c}}$). In the case of muonium formation, we attribute $\sigma_{{\rm c}}$ to an artifact resulting from an incomplete separation of $A_{\rm s}$ and $A_{\rm f}$ in the curve fit that might have led to residual depolarization, and we model the field dependence as follows: 
\begin{align}
	\sigma_{\rm s}(h)&=z\sigma_0(h_1)+(1-z)\sigma_0(h_2)+\sigma_{\rm c}\label{sgmh}\\
	\sigma_0(h_i)&=0.0274\times\frac{\gamma_{\mu}\Phi_0}{\lambda^2}(1-h_i)\left[1+3.9(1-h_i)^2\right]^{\frac{1}{2}},\label{brandt}
\end{align}
where $h_i\equiv H/H_{\rm c2}^{(i)}$, $H_{\rm c2}^{(2)}$ is the secondary upper critical field (a parameter corresponding to $\Delta_2$), $z$ is the relative weightf of $\sigma_{\rm s}(h_1)$, and Eq.~(\ref{brandt}) is an approximated expression for the field dependence of \sgm\ for a single-gap case (with the Ginzburg-Landau parameter $\kappa\gg1$) \cite{Brandt:88}.
 In the curve fit, $H_{\rm c2}^{(1)}$ was fixed to the reported value (5~T, determined by the specific heat \cite{Taguchi:05}). This model reproduces our data excellently, yielding $z=0.22(3)$ and $H_{\rm c2}^{(2)}=1.2(1)$ T.
A good agreement between the ratios $H_{\rm c2}^{(1)}/H_{\rm c2}^{(2)}=4.2(2)$ and $(\Delta_1/\Delta_2)^2=4.1(5)$ is perfectly in line with the general relation $H_{\rm c2}\propto\Delta^2$. (Note that the ratio $\Delta_1/\Delta_2$ is also in excellent agreement with that evaluated from the $T$ dependence.) More interestingly, the value of $H_{\rm c2}$ corresponds to the field toward which $\gamma$ exhibits a steep increase and then gradually saturates to $\gamma_n$ for $H_{\rm c2}^{(2)}\le H\le H_{\rm c2}^{(1)}$ \cite{Taguchi:05}. This, together with the $T$ dependence of \sgm, strongly suggests that the superconducting order parameter is highly anisotropic in the sample with $x=0.12$ that is characterized by a secondary energy scale $\Delta_2\simeq0.5\Delta_1$, where $\Delta_2$ may result from dips in a single-band gap or double-gap structure. Here, it would be worth quoting a prediction of band-structure calculation that carrier filling to the secondary band would not occur until $x$ exceeds $\sim$0.3, where a large increase in $N(0)$ is expected \cite{Heid, Felser:99}. Absence of such an increase in the recent specific heat measurement over the relevant range of $x$ supports the one-band scenario \cite{Kasahara:09}, suggesting that $\Delta_2$ corresponds to the minimal gap energy (at the dips) in the order parameter.

According to a recent theory based on the Hubbard model considering disconnected Fermi surfaces on a honeycomb lattice \cite{kuroki:08}, 
spin fluctuation enhances $T_{\rm c}$ over a low $x$ region. Since spin fluctuation-mediated superconductivity needs to have sign change in the gap function, the structure of the relevant Fermi surface leads to the prediction of $d+id'$ pairing symmetry as the most probable candidate. The decrease in $T_{\rm c}$ with an increase in doping is explained by an increase in three dimensionality and associated reduction in the relative volume in the Brillouin zone, where the pairing interaction is strong (corresponding to wave vectors that bridge the opposite sides of each pieces of the Fermi surface). Moreover, the anisotropy attributed to the dips in the order parameter develops for $x\ge0.11$ and that the minimum of gap energy is reduced to $\Delta_{\rm min}\sim0.4\Delta_{\rm max}$ for $x=0.16$ \cite{kuroki:09}. This behavior is in qualitative agreement with the tendency suggested by the $x$ dependence of the gap ratio $2\overline{\Delta}/k_{\rm B}T_{\rm c}$ deduced from the $T$ dependence of \sgm, where the secondary energy scale suggested in the case of $x=0.12$ ($\Delta_2\simeq0.5\Delta_1$) may correspond to the minimal gap ($\Delta_{\rm min}\sim\Delta_2$). Then, it is likely that the decrease in the gap ratio with an increase in $x$ reflects a decrease in $\Delta_{\rm min}$ in the $d+id'$ gap. 

\begin{figure}[ht]
	\centering
	\vspace{2mm}
	\includegraphics[width=0.7\linewidth,clip]{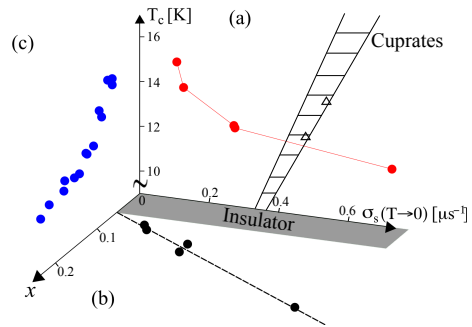}
	\caption{(Color online) (a) $T_{\rm c}$ as a function of \sgm$(T\rightarrow0)$. Solid circles show the present data, and open triangles are quoted from \cite{Ito:04} (only for Li$_x$ZrNCl with $x=0.17$ and 0.4, where the latter corresponds to the point with lower $T_{\rm c}$). The hatched area represents the empirical linear relation found in underdoped cuprates \cite{Ito:04}. (b) $x$ dependence of \sgm\ (the present data) with the dashed line representing a linear relation. This line crosses zero at $x\simeq0.05$, where metal-insulator transition occurs. (c) $x$ dependence of $T_{\rm c}$ (after Ref.\cite{Taguchi:06}).} 
	\label{tcsgm}
\end{figure}

Finally, we discuss the behavior of $\sigma_{\rm s}(T\rightarrow0)$ as a function of $T_{\rm c}$. As shown in Fig.~\ref{tcsgm}(a), $T_{\rm c}$ increases with a decrease in \sgm. It is inferred from our data shown in Fig.~\ref{tcsgm}(b) that \sgm\ is mostly proportional to $x$. Thus, the $\sigma_{\rm s}(x)$ dependence of $T_c$ is in excellent agreement with the earlier result that is shown in Fig.~\ref{tcsgm}(c) \cite{Taguchi:06}, which strongly supports the high quality of the investigated samples. Meanwhile, this is in marked contrast with the \sgm\ dependence of $T_{\rm c}$ reported in an earlier literature \cite{Ito:04}, where the authors maintain a linear relation common to that observed in underdoped cuprate superconductors (indicated by shaded area) based on their result obtained for $x=0.17$ and 0.4 (open triangles). In this regard, we point out the fact that the Li$_x$ZrNCl samples studied in Ref.~\cite{Ito:04} do not follow the $T_c$ versus $x$ relation observed in our samples. In particular, their sample with $x=0.17$ seems to exhibit relatively high $T_{\rm c}$ corresponding to that of $x=0.07$--0.08 in our sample. Considering the difficulty associated with obtaining a uniform specimen for a low Li concentration \cite{Taguchi:06}, one might speculate that their sample (particularly for $x=0.17$) might have had problems concerning homogeneity. The reported small \msr\ asymmetry ($\simeq$0.1 \cite{Ito:04}) might be further evidence for this speculation. In any case, our result shows that the $n_{\rm s}$ dependence of $T_c$ is markedly different from that observed in the case of underdoped high $T_{\rm c}$ cuprates. 

In conclusion, it is microscopically shown by \msr\ measurements that the superconducting order parameter in Li$_x$ZrNCl is strongly anisotropic in a sample with $x=0.12$, as inferred from the temperature and field dependence of \sgm. Therefore, the origin of anomalous behavior in the Sommerfeld coefficient observed in the specific heat measurements is attributed to the strong anisotropy in the order parameter that might be characterized by a secondary energy scale $\Delta_2$ (where $\Delta_2\simeq0.5\Delta_1$). The temperature dependence of \sgm\ over a range of $x$ from 0.10 to 0.21 implies a tendency for the occurrence of a weak pairing interaction and strong anisotropy in the order parameter with an increase in $x$. The latter feature as well as the relatively high $T_{\rm c}$ at a low Li concentration supports the important role of electronic correlation for the mechanism of superconductivity in Li$_x$ZrNCl predicted by the recent theory. 

We thank the staff of TRIUMF for their technical support during the \msr\ experiment, and we appreciate the helpful discussion with Profs. K. Kuroki and H. Tou. This work was supported by the KEK-MSL Inter-University Program for Oversea Muon Facilities and a Grant-in-Aid for Scientific Research on Priority Areas by Ministry of Education, Culture, Sports, Science and Technology, Japan.

% Produces the bibliography via BibTeX.

\begin{thebibliography}{00}
	\bibitem{Yamanaka:96}
		S.~Yamanaka, H.~Kawaji, K.~Hotehama, and M.~Ohashi, Adv. Mater. {\bf 8}, 771 (1996).
	\bibitem{Yamanaka:98} 
		S.~Yamanaka, K.~Hotehama, and H.~Kawaji, Nature {\bf 392}, 580 (1998).
	\bibitem{Taguchi:05}
		Y.~Taguchi, M.~Hisakabe, and Y.~Iwasa, Phys. Rev. Lett. {\bf 94}, 217002 (2005).
	\bibitem{Tou:01}
		H.~Tou, Y.~Maniwa, T.~Koiwasaki, and S.~Yamanaka, Phys. Rev. Lett. {\bf 86}, 5775 (2001).
	\bibitem{Weht}
		R.~Weht, A.~Filippetti, and W.~E.~Pickett, Europhys. Lett. {\bf 48}, 320 (1999).
	\bibitem{Heid}
		R.~Heid and K.-P.~Bohnen Phys. Rev. B {\bf 72} 134527 (2005).
	\bibitem{Taguchi:06}
		Y.~Taguchi, A.~Kitora, and Y.~Iwasa, Phys. Rev. Lett. {\bf 97}, 107001 (2006).
	\bibitem{kuroki:08}
		K.~Kuroki, Sci. Technol. Adv. Mater. {\bf 9} 044202 (2008).
	\bibitem{kuroki:09}
		K.~Kuroki, arXiv: 1001.3167.
	\bibitem{Takano:08} T. Takano, A. Kitora, Y. Taguchi, and Y. Iwasa,
		Phys. Rev. B {\bf 77}, 104518 (2008).
	\bibitem{Bouqyuet:01}
		F.~Bouquet, Y.~Wang, R.~A.~Fisher, D.~G.~Hinks, J.~D.~Jorgensen, A.~Junod, and N.~E.~Phillips: Europhys.~Lett. {\bf56} (2001) 856.
	\bibitem{Ohishi:01}
		K.~Ohishi, T.~Muranaka, J.~Akimitsu, A.~Koda, W.~Higemoto, and R.~Kadono, J. Phys. Soc. Jpn. {\bf 72}, 29 (2003).
	\bibitem{Brandt:88}
		E.~H.~Brandt, Phys. Rev. B {\bf 37}, 2349 (1988).
	\bibitem{Felser:99} 
		C. Felser and R. Seshadri, J. Mater. Chem. {\bf 9}, 459 (1999).
	\bibitem{Kasahara:09} 
		Y. Kasahara, T.~Kishiume, T.~Takano, K.~Kobayashi, E.~Matsuoka, H.~Onodera, K.~Kuroki, Y.~Taguchi, and Y.~Iwasa, Phys. Rev. Lett. {\bf 103} 077004 (2009).
	\bibitem{Ito:04}
		T.~Ito, Y.~Fudamoto, A.~Fukaya, I.~M.~Gat-Malureanu, M.~I.~Larkin, P.~L.~Russo, A.~Savici, Y.~J.~Uemura, K.~Groves, R.~Breslow, K.~Hotehama, S.~Yamanaka, P.~Kyriakou, M.~Rovers, G.~M.~Luke, and K.~M.~Kojima, Phys. Rev. B {\bf 69}, 134522 (2004).
\end{thebibliography}
\end{document}